\documentclass[useAMS,usenatbib]{mn2e}
\usepackage{graphicx}
\usepackage{color}
\usepackage{soul}
\usepackage{bm}

\newcommand{\bec}[1]{\mbox{\boldmath $ #1$}}

\newcommand{\meanT}{\overline{T}}
\newcommand{\meanrho}{\overline{\rho}}
\newcommand{\meanIb}{\overline{I}_b}
\newcommand{\meanI}{\overline{I}}
\newcommand{\meankappa}{\overline{\kappa}}

\title[Turbulent transport of radiation in the solar convective zone]{Turbulent transport of radiation in the solar convective zone}

\author[I. Rogachevskii, N. Kleeorin]
{I. Rogachevskii,
 N. Kleeorin\\
   Department of Mechanical Engineering,
        Ben-Gurion University of Negev, POB 653, 8410530 Beer-Sheva, Israel\\
   Nordita, KTH Royal Institute of Technology and Stockholm University,
        Hannes Alfvens 12, 10691 Stockholm, Sweden}

\begin{document}

%\pagerange{\pageref{firstpage}--\pageref{lastpage}}
%\pubyear{2015}

\maketitle

%\label{firstpage}

\begin{abstract}
A turbulent transport of radiation in the solar convective zone is investigated.
The mean-field equation for the irradiation intensity is derived.
It is shown that due to the turbulent effects, the effective penetration length of radiation can be increased in several times in comparison with the mean penetration length of radiation (defined as an inverse mean absorption coefficient). Using the model of the solar convective zone based on the mixing length theory, where the mean penetration length of radiation is usually much smaller than the turbulent correlation length, it is demonstrated that the ratio of the effective penetration length to the mean penetration length of radiation increases in 2.5 times in the vicinity of the solar surface.
The main reason are the compressibility effects that become important in the vicinity of the solar surface where temperature and density fluctuations increase towards the solar surface,
enhancing fluctuations of the radiation absorption coefficient and increasing the effective penetration length of radiation.
\end{abstract}

%Preprint NORDITA 2021-074

\maketitle


\section{Introduction}

Turbulent transport of temperature, particles and magnetic fields
have been studied analytically, in laboratory and field experiments
and in numerical simulations for more than a century \citep[see, e.g.,][]{MY71,MY75,MC90,F95,LE08,DA13,RI21},
however, some fundamental questions remain.
This is particularly true in applications to astrophysics,
where the governing parameter values are too
extreme to be modelled either experimentally or numerically.

In astrophysical turbulent flows radiative transport can be affected by
turbulence.
This effect is different for optically
thick and optically thin regimes of the radiative transport.
For optically thick regime, the mean free path of photons is much smaller
in comparison with the typical scales of the flow.
In the opposite case, i.e., for optically thin regime, the mean free path of photons is much
larger than the fluid motion scales, so the photons propagate over
large distances before they are absorbed and re-emitted
again \citep[see, e.g.,][]{CH60,MM84,AK96,LI02,HM10}.

The relaxation time of small temperature perturbations by radiative diffusion
has been determined by \cite{SP57}, where
a time-dependent equation for the temperature field of a medium with deviations from radiative
equilibrium has been derived assuming that the medium is graylike and there are no internal motions or compressional effects, and heat is exchanged only radiatively.
It has been shown that perturbations of small amplitude imposed on a homogeneous medium decay exponentially in time, and the decay time depends on a characteristic
length of the perturbations \citep{SP57}.

Turbulent diffusion can be increased by a radiative diffusion, i.e., by a photon diffusion.
For instance, the decay rates of sinusoidal large-scale temperature
perturbations in the optically thick and thin regimes have been
determined by \cite{BD21} using radiative hydrodynamic direct numerical simulations
of forced turbulence. It has been shown there that the rate of decay increases
with the wavenumber. However, this effect is much weaker in comparison with the
effect of the standard turbulent diffusion \citep{BD21}.

In the present study, we investigate the turbulent transport of radiation in the
solar convective zone. We derive a mean-field equation for the irradiation intensity, and
show that the effective penetration length of radiation can be increased by turbulence
in several times in comparison with the mean penetration length of radiation
which is defined as an inverse mean absorption coefficient.
This effect has been tested using the model of the solar convective zone \citep{SP74} based on the mixing length theory. According to this model, the mean penetration length of radiation is much less than the turbulent correlation length.
We show that the ratio of the effective penetration length to the mean penetration length of radiation increases in 2.5 times in the vicinity of the solar surface.

This paper is organized as follows.
In Section~2, we discuss a general concept of turbulent transport
of radiation,
and in Section~3 we derive the mean-field radiation transport equation.
In Section~4, we derive expression for the effective penetration
length of radiation in turbulent flows, which depends on
the ratio of fluctuations of the radiation absorption
coefficient to the mean penetration
length of radiation.
To calculate the effective penetration
length of radiation, in Section~4 we determine
fluctuations of the radiation absorption
coefficient which are caused by fluctuations of fluid temperature
and density, and the temperature-density correlations.
In Section~5, we apply obtained results to the solar convective zone.
Finally, in Section~6, we discuss our results and draw conclusions.
In Appendix~A we derive expressions for the level of temperature and density fluctuations
as well as temperature--density correlations which allow us to determine
fluctuations of the radiation absorption coefficient.

\section{General concept of turbulent transport of radiation}

In solar and stellar convective zones, the convective transport of energy
is more effective than the radiative transfer.
The Schwarzschild criterion for the onset of convection
does not take into account an effect of turbulence
on radiative transfer.
However, the absorption coefficient of radiation
depends on temperature and density, and
there are strong fluctuations of
the fluid temperature and density
in solar and stellar convective zones.
These fluctuations affect the absorption coefficient
of radiation, and therefore they
affect turbulent transfer of radiation.

In solar and stellar convective zones,
the characteristic times of turbulent motions are much larger than the
radiation time, and the integral turbulent scales are much larger
than the mean penetration length of
radiation defined as the inverse absorption coefficient of radiation.
The latter implies that turbulent eddies are optically thick,
and inhomogeneities in the fluid temperature and density
can strongly affect the radiation transfer.

To describe the radiation transfer, we use
the radiation transport equation.
This equation represents a steady state version of the
equation for electromagnetic energy transfer,
since the time of photon propagation
(the radiation time) is very short.
This equation is characterized by the absorption coefficient
of radiation and the black-body radiation
intensity of the gas.
The radiative transport equation for the intensity $I({\bm r},\hat{\bm s},\omega)$
reads \citep[see, e.g.,][]{CH60,MM84,AK96,LI02,HM10}:
\begin{eqnarray}
&& \left(\hat{\bm s} {\bm \cdot} {\bm \nabla}\right)
I({\bm r},\hat{\bm s},\omega) = - \kappa({\bm r},\omega) \left[I - I_b(T,\omega)\right] ,
\label{LA4}
\end{eqnarray}
where ${\bm r}$ is the position vector,
$\hat{\bm s}={\bm k}/k$ is the unit vector in the direction of radiation,
${\bm k}$ is the wave vector, $\kappa({\bm r},\omega)= \rho \, \kappa_{\rm op}$ is the absorption coefficient of gas,
$\kappa_{\rm op}=\kappa_0 \, \rho^a \, T^b$ is the opacity of the gas, $T$ and $\rho$
are the gas temperature and density,
$I_{b}(T,\omega)$ is the black-body radiation intensity of the gas,
and $\omega$ is the radiation frequency.
Here we take into account the radiation absorption in gases and neglect
the radiation scattering in gases.
The function $I_{b}(T,\omega)$ in a local thermodynamic equilibrium is given by
\begin{eqnarray}
I_{b}(T,\omega) = {\hbar \omega^3 \over \pi^2 c^3} \,
\left[\exp \left({\hbar \omega \over k_{\rm B} T} \right) - 1 \right]^{-1} ,
\label{LAAA4}
\end{eqnarray}
where $\hbar$ is Planck's constant, $c$ is the speed of light
and $k_{\rm B}$ is the Boltzman constant.
The integral $\int I_{b}(T,\omega) \, d \omega \propto \sigma T^4$
yields the Stefan-Boltzmann law.

Our goal is to derive effective radiation transport equation
with effective transport coefficients: effective absorption coefficient of radiation
and the effective source of the radiation intensity.
To take into account the turbulence effects,
we apply a mean-field approach and average
the radiation transport equation~(\ref{LA4}) over ensemble
of fluctuations. In the framework of the mean-field approach, all
quantities are decomposed into the mean and fluctuating parts:
$I=\meanI + I'$, $I_b=\meanIb + I'_b$ and $\kappa=\meankappa + \kappa'$.
We adopt the Reynolds averaging, where $\meanI=\langle I \rangle$, $\meanIb=\langle I_b \rangle$, $\meankappa=\langle \kappa \rangle$ are the mean fields, and $I', I'_b, \kappa'$ are the fluctuating fields with zero mean, and the angular brackets denote ensemble averaging.
To derive the mean-field radiation transport equation, we adopt a method applied by \cite{KK89,LKRH17,LKRH18}.

The obtained mean-field equation contains
the correlation function for fluctuations of
the absorption coefficient of radiation $\kappa'$ and
the radiation intensity $I'$, i.e.,
$\langle \kappa' \, I' \rangle$.
This correlation is due to fluctuations of
temperature and density.
This equation also contains the correlation function for fluctuations of
the absorption coefficient of radiation $\kappa'$ and
the black-body radiation intensity of the gas
$\langle \kappa' \, I'_b \rangle$ due to fluctuations of
temperature.

To determine the correlation functions, $\langle \kappa' \, I' \rangle$
and $\langle \kappa' \, I'_b \rangle$, we derive equation for
fluctuations of the radiation intensity $I'$ by
subtracting the obtained mean-field equation from the radiation
transport equation~(\ref{LA4}).
Since the equation for fluctuations of the radiation intensity $I'$
is a linear equation, we solve this equation exactly.
However, the solution of this equation is nonlinear in
fluctuations of $\kappa'$. This causes appearance
of the high-order moments in fluctuations of $\kappa'$ in the expression
for the correlation function $\langle \kappa' \, I' \rangle$.
We assume that fluctuations of $\kappa'$ are essentially less
than the mean absorption coefficient of radiation.
This allows us to obtain the closed results.

The main expected result of this study is that
the derived mean-field equation
for the radiation transfer with the effective transport coefficients yields
the effective penetration length of radiation.
When the effective penetration length of radiation is larger than
the mean penetration length of radiation, the absorbtion coefficient decreases and an observer
can see more deeper layers inside the stars.
The reasons for the increase of the effective penetration length of radiation
in turbulent flows are caused by the correlation between fluctuations of
the radiation absorption coefficient $\kappa'$ and fluctuations of
the irradiation intensity $I'$.
We show below that this correlation function should be
negative, because an increase of the absorption
of radiation decreases the radiation intensity and wise versa.
We also demonstrate in this study that this effect is essential in the vicinity
of the solar surface.

\section{Mean-field radiation transport equation}  	

In this section we derive the mean-field radiation transport equation.
Ensemble averaging of equation~(\ref{LA4}) yields
the equation for the mean radiation intensity $\meanI$:
\begin{eqnarray}
&& \left(\hat{\bm s} {\bm \cdot} {\bm \nabla}\right)\meanI = - \meankappa
\left(\meanI - \meanIb\right) - \langle \kappa' \, I' \rangle
+ \langle \kappa' \, I'_b \rangle .
\label{LA5}
\end{eqnarray}
This equation contains unknown correlation functions, $\langle \kappa' \, I' \rangle$
and $\langle \kappa' \, I'_b \rangle$.
To determine these correlation functions,
we derive equation for fluctuations of the radiation intensity $I'$.
To this end, we subtract the mean-field radiation transport equation~(\ref{LA5})
from equation~(\ref{LA4}), so that the equation for fluctuations of $I'$ reads:
\begin{eqnarray}
&& \left(\hat{\bm s} {\bm \cdot} {\bm \nabla} + \meankappa
+ \kappa' \right) I'({\bm r},\hat{\bm s}) = I_{\rm source} ,
\label{LA6}
\end{eqnarray}
where the source term $I_{\rm source}$ is given by
\begin{eqnarray}
I_{\rm source} &=& - \kappa' \, \left(\meanI - \meanIb\right)
+ \langle \kappa' \, I' \rangle +
\left(\meankappa + \kappa'\right)I'_b
- \langle \kappa' \, I'_b \rangle  .
\nonumber\\
\label{LA7}
\end{eqnarray}
The solution of equation~(\ref{LA6}) reads
\begin{eqnarray}
I'({\bm r},\hat{\bm s}) &=& \int_{-\infty}^{\infty}
\exp\left[-\left|\int_{s'}^{s}
\left[\meankappa+ \kappa'(s'') \right] \,ds'' \right| \right]
\nonumber\\
&& \times I_{\rm source}(s')  \,ds' ,
\label{LA8}
\end{eqnarray}
where $s={\bm r} {\bm \cdot} \hat{\bm s}$.
This solution is nonlinear in fluctuations of $\kappa'$.
The latter causes appearance
of the high-order moments in fluctuations of $\kappa'$ in the expression
for the correlation function $\langle \kappa' \, I' \rangle$.
The high-order moments are much less than the lower-order moments, because
$\kappa' \ll \meankappa$. This allows us to
expand the function, $\exp\left[-\int_{s'}^{s} \kappa'(s'') \,ds''\right]$, in equation~(\ref{LA8}) in Taylor series:
\begin{eqnarray}
\exp\left[-\int_{s'}^{s} \kappa'(s'') \,ds''\right] = 1 - \int_{s'}^{s} \kappa'(s'') \,ds'' + {\rm O}\left(\kappa'^2\right) .
\nonumber\\
\label{LA8a}
\end{eqnarray}
Therefore, equation~(\ref{LA8}) can be rewritten as:
\begin{eqnarray}
I'({\bm r},\hat{\bm s}) &=& \int_{-\infty}^{\infty} I_{\rm source}(s') \,
\exp\left(-\meankappa |s-s'|\right)
\nonumber\\
&& \times \left(1 - \int_{s'}^{s} \kappa'(s'') \,ds'' \right) \,ds'
+ {\rm O}\left(\kappa'^2\right) .
\label{LA8b}
\end{eqnarray}
Multiplying equation~(\ref{LA8b}) by $\kappa'$ and averaging over the ensemble, we obtain expression for the one-point correlation function $\langle \kappa' \, I' \rangle$:
\begin{eqnarray}
&& \langle \kappa' \, I' \rangle \, \biggl[1 + \int_{-\infty}^{\infty}
\biggl(\int_{s'}^{s} \langle\kappa'(s) \kappa'(s'') \rangle \,ds'' \biggr)
\nonumber\\
&& \quad \times \exp\biggl(-\meankappa |s-s'|\biggr) \, \,ds' \biggr] = -  \biggl[ \int_{-\infty}^{\infty} \langle\kappa'(s) \kappa'(s') \rangle
\nonumber\\
&& \quad \times \exp\biggl(-\meankappa |s-s'|\biggr) \, \,ds'\biggr] \, \Big(\meanI - \meanIb\Big),
\label{LA8c}
\end{eqnarray}
where we neglect the third-order and higher-order moments in fluctuations of $\kappa'$.
Equation~(\ref{LA8c}) can be rewritten as
\begin{eqnarray}
\langle \kappa' \, I' \rangle  = - \meankappa
\Big(\meanI - \meanIb\Big) \, {2 \meankappa J_1 \over 1 + 2 \meankappa J_2} ,
\label{LA9}
\end{eqnarray}
where the integrals $J_1$ and $J_2$  in equation~(\ref{LA9}) are defined as:
\begin{eqnarray}
&& J_1= \int_{0}^{\infty} \Phi(Z) \exp(-\meankappa Z) \,dZ ,
\label{LA10}\\
&& J_2= \meankappa \int_{0}^{\infty} \left(\int_{0}^{Z} \Phi(Z') \,dZ' \right)
\, \exp(-\meankappa Z) \,dZ ,
\label{LA11}
\end{eqnarray}
the function $\Phi(Z)$ is defined as $\Phi(Z)=\langle\kappa'(s) \kappa'(s') \rangle$ and $Z=|s-s'|$.

Substituting equation~(\ref{LA9}) into the mean-field equation~(\ref{LA5}), we arrive at
the mean-field radiation transport equation as
\begin{eqnarray}
&& \left(\hat{\bm s} {\bm \cdot} {\bm \nabla}\right)\meanI = - \kappa_{\rm eff} \,
\left(\meanI - I_{b}^{\rm eff}\right) ,
\label{LA12}
\end{eqnarray}
where the effective absorption coefficient $\kappa_{\rm eff}$ is given by
\begin{eqnarray}
\kappa_{\rm eff} = \meankappa \, \left(1- {2 \meankappa J_1 \over 1 + 2 \meankappa J_2}\right) ,
\label{LA14}
\end{eqnarray}
and the effective radiation intensity is
\begin{eqnarray}
I_{b}^{\rm eff} =\meanIb + {\langle \kappa' \, I'_b \rangle \over \kappa_{\rm eff}} .
\label{LA15}
\end{eqnarray}
The function $\meanIb$ is expanded in Taylor series as
\begin{eqnarray}
\meanIb = \left[I_{b} + {\langle\theta^2 \rangle \over 2} \, {\partial^2 I_{b} \over \partial T^2} \right]_{T=\meanT} ,
\label{LA16}
\end{eqnarray}
where the fluid temperature is
decomposed into the mean $\meanT$ and fluctuating $\theta$ parts:
$T= \overline{T} + \theta$.
Solution of the mean-field radiation transport equation~(\ref{LA12}) for the mean
irradiation intensity $\meanI({\bm r},\hat{\bm s},\omega)$ is given by
\begin{eqnarray}
&& \meanI({\bm r},\hat{\bm s},\omega) = \int_{-\infty}^{\infty}
I_{b}^{\rm eff}({\bm r}',\omega) \, \exp\left[-\left|\tau({\bm r},{\bm r}',\hat{\bm s}) \right| \right]\, \hat{\bm s} {\bm \cdot} \,d{\bm r}' ,
\nonumber\\
\label{LA17}
\end{eqnarray}
where $\tau({\bm r},{\bm r}',\hat{\bm s}) = \int_{{\bm r}}^{{\bm r}'} \kappa_{\rm eff}({\bm r}'') \, \hat{\bm s} {\bm \cdot} \,d{\bm r}''$ is the optical depth.

\section{Effective penetration length of radiation and fluctuations of absorption coefficient}

In this section we determine the effective penetration
length of radiation in turbulent flows, defined as
$L_{\rm eff} = \kappa_{\rm eff}^{-1}$.
Since the main contribution to the second moment $\langle\kappa'(s) \kappa'(s') \rangle$ of
fluctuations of the absorption coefficient
is from the integral scale of turbulence $\ell_0$,
we assume that this correlation functions have the form:
\begin{eqnarray}
\langle\kappa'(s) \kappa'(s') \rangle = \left\langle \kappa'^{\, 2} \right\rangle \, \exp\left(-{|s-s'| \over \ell_0}\right).
\label{LB4}
\end{eqnarray}
Using equations~(\ref{LA10})--(\ref{LA11}) and~(\ref{LB4}), we calculate the integrals $J_1$ and $J_2$ as
\begin{eqnarray}
J_1=J_2 = {\left\langle \kappa'^{\, 2} \right\rangle \over \meankappa^3} \, \left(1+ {L_r \over \ell_0}\right)^{-1} ,
\label{LB5}
\end{eqnarray}
where $L_{r} = \meankappa^{\,-1}$ characterises the mean penetration
length of radiation in turbulent flow.
Therefore, equations~(\ref{LA14}) and~(\ref{LB5}) allow us to determine
the effective penetration length $L_{\rm eff} = \kappa_{\rm eff}^{-1}$
of radiation in a turbulent flow as
\begin{eqnarray}
L_{\rm eff} = L_r \, \left[1 + {2 \, \left\langle \kappa'^{\, 2} \right\rangle \over \meankappa^2} \, \left(1+ {L_r \over \ell_0}\right)^{-1} \right] .
\label{LB6}
\end{eqnarray}
We consider two limiting cases:

(i) $\ell_0 \ll L_r$, the effective penetration length $L_{\rm eff}$ is
\begin{eqnarray}
L_{\rm eff} = L_r \, \left[1 + {2 \, \left\langle \kappa'^{\, 2} \right\rangle \over \meankappa^2} \, {\ell_0 \over L_r}\right] ,
\label{LB7}
\end{eqnarray}

(ii) $\ell_0 \gg L_r$, the effective penetration length $L_{\rm eff}$ is
\begin{eqnarray}
L_{\rm eff} = L_r \, \left[1 + {2 \, \left\langle \kappa'^{\, 2} \right\rangle \over \meankappa^2}
\right] .
\label{LB8}
\end{eqnarray}
Equation~(\ref{LB8}) implies that for $\ell_0 \gg L_r$, the effective penetration length $L_{\rm eff}$ can increase in 3 times in comparison with the mean penetration
length $L_r$ of radiation due to the turbulence effects when $\left\langle \kappa'^{\, 2} \right\rangle \sim \meankappa^2$.

The mechanism of increase of the effective penetration length $L_{\rm eff}$
in turbulent flows is related to the correlation between fluctuations of
the radiation absorption coefficient $\kappa'$ and fluctuations of
the irradiation intensity $I'$. The correlation $\langle \kappa' \, I' \rangle$
is negative because an increase of the absorption of radiation decreases
the radiation intensity and wise versa. Fluctuations of
the radiation absorption coefficient are caused by fluctuations of
fluid temperature and density in turbulent flow.

Now we determine fluctuations of
the radiation absorption coefficient.
For simplicity, we assume that the opacity of gas is $\kappa_{\rm op}=\kappa_0 \, \rho^a \, T^b$, so that the absorption coefficient of gas is $\kappa= \rho \, \kappa_{\rm op}=\kappa_0 \, \rho^{a+1} \, T^b$. According to the Schwarzschild stability criterion, the case $a=1$ and $b=0$ corresponds to
the marginally stable regime, while the case $a=1$ and $b=1$ corresponds to the unstable regime
\citep{BB14}.
The equation $\kappa= \kappa_0 \, \rho^{a+1} \, T^b$ allows us to determine the ratio of fluctuations of the absorption coefficient $\kappa'$ to the mean value of $\meankappa$ as
\begin{eqnarray}
{\kappa' \over \meankappa} = (a+1) \, {\rho' \over \meanrho} + b \, {\theta \over \meanT} ,
\label{LB1}
\end{eqnarray}
where $\rho'$ are density fluctuations and $\meanrho$ is the mean fluid density.

Using equation~(\ref{LB1}), we determine the correlation function $\langle \kappa' \, I'_b \rangle$  as:
\begin{eqnarray}
\langle \kappa' \, I'_b \rangle = \meankappa  \,\left[(a+1) \, {\langle\rho' \theta\rangle\over \meanrho} + b \, {\langle\theta^2 \rangle\over \meanT} \right] \, \left({\partial I_{b} \over \partial T}\right)_{T=\meanT} ,
\label{LC1}
\end{eqnarray}
where we take into account that $I'_b = \theta \, (\partial I_{b}/ \partial T)_{T=\meanT}$.
To find the effective penetration
length of radiation, we determine the level of fluctuations $\left\langle \kappa'^{\, 2} \right\rangle$
of the absorption coefficient as
\begin{eqnarray}
{\left\langle \kappa'^{\, 2} \right\rangle \over \meankappa^2} &=& (a+1)^2 \, {\left\langle \rho'^{\, 2} \right\rangle \over \meanrho^2}
+ b^2 \,{\left\langle \theta^2 \right\rangle \over \meanT^2}
+ 2 b \, (a+1) \, {\left\langle \theta \, \rho'\right\rangle \over \meanrho \, \meanT} .
\nonumber\\
\label{LC2}
\end{eqnarray}
The intensity of temperature fluctuations in a developed compressible turbulence for large P\'eclet and Reynolds numbers is given by
\begin{eqnarray}
{\left\langle  \theta^2 \right\rangle \over \overline{T}^2}&=& 8 \, f_c \, (\gamma-1)^2  \left({\sigma_c \over 1 + \sigma_c} \right)^3 \, \biggl[1 - {(\lambda \, \ell_0)^2\over 9} \biggr]
\nonumber\\
&& +  {8 \over 9}  \, \ell_0^2 \, \biggl[{{\bm \nabla}\overline{T} \over \overline{T}} + (\gamma-1) \, {\bm \lambda}\biggr]^2 ,
\label{DD1}
\end{eqnarray}
(see Appendix~A), where $\gamma=c_{\rm p}/c_{\rm v}$ is the ratio of specific heats,
${\bm \lambda} = -{\bm \nabla} \ln \overline{\rho}$ characterizes the inhomogeneity
of the mean fluid density, the parameter
\begin{eqnarray}
\sigma_c = {\left\langle (\bec{\nabla} \cdot \, {\bm u})^2 \right\rangle
\over \left\langle(\bec{\nabla} \times {\bm u})^{2} \right\rangle}
  \label{APC44}
\end{eqnarray}
is the degree of compressibility of the turbulent velocity field ${\bm u}$
and $\ell_0$ is the integral scale of turbulence.
The function $f_c(q,q_c,\sigma_c)$ depends on the degree of compressibility and the exponents of spectra $q$ and $q_c$ for the incompressible and compressible parts of velocity fluctuations (see Appendix~A):
\begin{eqnarray}
f_c &=& {q_c -1 \over 3 q_c -5} + {2 (q_c -1) \over \sigma_c (q + 2 q_c -5)} + {q_c -1 \over \sigma_c^2 (2 q + q_c -5)} .
\nonumber\\
\label{C5}
\end{eqnarray}
Equation~(\ref{DD1}) is different from that derived by \cite{RK21}.
In this study, we take into account a strong density stratification.
The latter is important in view of applications to the solar convective zone,
where the fluid density in radial direction is changed by seven orders of magnitude.
We also neglect the gradient of the turbulent diffusion that
is changed very slowly in the solar convective zone.
The first term in the right hand side of equation~(\ref{DD1}) determines
a compressibility contribution of velocity fluctuations
to temperature fluctuations, while the second term in equation~(\ref{DD1})
is proportional to the squared gradient of the mean entropy.

The level of fluid density fluctuations in a developed compressible turbulence for large P\'eclet and Reynolds numbers is given by
\begin{eqnarray}
&& {\left\langle \rho'^{\, 2} \right\rangle \over \meanrho^2}
 = 8 \, f_c \, \left({\sigma_c \over 1 + \sigma_c} \right)^3 \, \biggl[1 - {(\lambda \, \ell_0)^2\over 9} \biggr] ,
\label{C8}
\end{eqnarray}
and the cross correlations $\left\langle \theta \, \rho'\right\rangle$ is
\begin{eqnarray}
{\left\langle \theta \, \rho'\right\rangle \over \meanrho \, \overline{T}}
&=& 8 \, f_c \, (\gamma-1) \, \left({\sigma_c \over 1 + \sigma_c} \right)^3 \, \biggl[1 - {(\lambda \, \ell_0)^2\over 9} \biggr] .
\label{C9}
\end{eqnarray}
Equations~(\ref{DD1}), (\ref{C8}) and~(\ref{C9}) are valid for small $\sigma_c < 1$. The latter condition is typical for developed turbulence and turbulent convection.
Derivation of equations~(\ref{DD1}), (\ref{C8}) and~(\ref{C9}) is given in Appendix~A.

Therefore, the level of fluctuations $\left\langle \kappa'^{\, 2} \right\rangle$
of the absorption coefficient is
\begin{eqnarray}
&& {\left\langle \kappa'^{\, 2} \right\rangle \over \meankappa^2} =
8 \, f_c \, \left({\sigma_c \over 1 + \sigma_c} \right)^3 \, \biggl[1 - {(\lambda \, \ell_0)^2\over 9} \biggr] \, \biggl[a+1
\nonumber\\
&& \quad + b \,(\gamma-1)\biggr]^2 +  {8 \over 9}  \, \ell_0^2 \, b^2 \,\biggl[{{\bm \nabla}\overline{T} \over \overline{T}} + (\gamma-1) \, {\bm \lambda}\biggr]^2.
\label{C11}
\end{eqnarray}
For $a=1$ and $b=1$, the level of fluctuations $\left\langle \kappa'^{\, 2} \right\rangle$
of the absorption coefficient is given by
\begin{eqnarray}
&& {\left\langle \kappa'^{\, 2} \right\rangle \over \meankappa^2} =
8 \, f_c \, \left({\sigma_c \over 1 + \sigma_c} \right)^3 \, \biggl[1 - {(\lambda \, \ell_0)^2\over 9} \biggr] \, (\gamma+1)^2
\nonumber\\
&& \quad +  {8 \over 9}  \, \ell_0^2 \, b^2 \,\biggl[{{\bm \nabla}\overline{T} \over \overline{T}} + (\gamma-1) \, {\bm \lambda}\biggr]^2 .
\label{CCC11}
\end{eqnarray}
Note that for nearly isentropic flows where ${\bm \nabla} \ln \overline{T} \approx (\gamma-1) \,{\bm \nabla} \ln \overline{\rho}$, the last term in equations~(\ref{C11})--(\ref{CCC11}) is small.
This term is proportional to the gradient of the mean entropy ${\bm \nabla} \overline{S}=c_{\rm v} [{\bm \nabla} \ln \overline{T} - (\gamma-1) \,{\bm \nabla} \ln \overline{\rho}]$.

\section{Application to the solar convective zone}

\begin{figure}
\centering
\includegraphics[width=7.5cm]{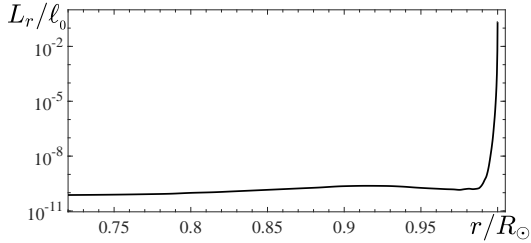}
\caption{\label{Fig1} The radial profile of the ratio $L_r / \ell_0$ for the solar convective zone.}
\end{figure}

\begin{figure}
\centering
\includegraphics[width=7.5cm]{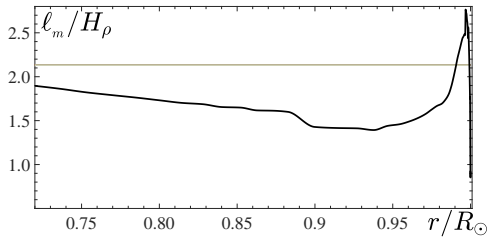}
\caption{\label{Fig2} The profile of the ratio  $\ell_{\rm m} / H_\rho$ of the mixing length $\ell_{\rm m}$ to the density stratification length $H_\rho$ versus $r/R_\odot$ that is based on the
model of the solar convective zone.}
\end{figure}

\begin{figure}
\centering
\includegraphics[width=7.0cm]{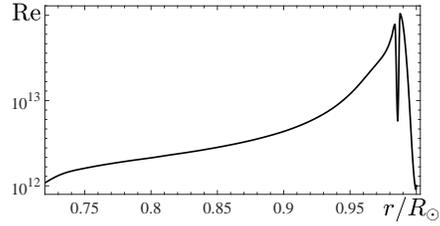}
\caption{\label{Fig3} The radial profile of the Reynolds number Re$(r) = u_0 \, \ell_0/\nu$ for the solar convective zone.}
\end{figure}

In this section, we apply the obtained results to the solar convective zone.
We use the model of the solar convective zone by \cite{SP74} based on the mixing length theory. According to this model, the mean penetration length of radiation is much less than the turbulent correlation length.
Indeed, in Figs.~\ref{Fig1}--\ref{Fig3} we show the radial profiles of the ratio of the mean penetration length of radiation to the integral scale of turbulence $L_r(r) / \ell_0$, the ratio  $\ell_{\rm m} / H_\rho$ of the mixing length $\ell_{\rm m}$ to the density stratification length $H_\rho=\lambda^{-1}$ and  the Reynolds number Re$(r) = u_0 \, \ell_0/\nu$ for the solar convective zone based on the model by \cite{SP74}.
The radius $r$ is measured in units of the solar radius $R_\odot$.
The mixing length $\ell_{\rm m}$ is identified with the size of the solar granulations, while
the ratio $\ell_{\rm m} /\ell_0 = 5 - 7$ is justified by the results of analytical study \citep{EKRZ02,EKRZ06} and laboratory experiments \citep{BKR09}, which show that the integral scale $\ell_0$ of the turbulent convection is smaller in 5 - 7 times in comparison with the size of the large-scale circulations.
These turbulent parameters increase towards the solar surface.

To determine the effective penetration length of radiation, we estimate
the degree of compressibility of the turbulent fluid flow for small Mach numbers as \citep{RK21}
\begin{eqnarray}
\sigma_c \sim {\rm Ma}^5 \, {\rm Re}^{1/4} ,
\label{M26}
\end{eqnarray}
where ${\rm Ma}=u_0/c_{\rm s}$ is the Mach number, $u_0=\left\langle {\bm u}^2\right\rangle^{1/2}$ and
$c_{\rm s} = (\gamma \overline{P}/\overline{\rho})^{1/2}$
is the sound speed, ${\rm Re}= u_0 \, \ell_0/\nu$
is the Reynolds number and $\nu$ is the kinematic viscosity.
The estimate~(\ref{M26}) is obtained assuming that the effect of compressibility to
the viscous heating $\overline{J}_\nu^{\,({\rm c})}$ is of the order of
the radiative wave energy density $E_{\rm w}$.
In particular, turbulence can generate acoustic waves, and
the rate of the energy radiated by the acoustic waves
per unit mass for small Mach numbers is given by \citep{L52,L54,P52}
\begin{eqnarray}
E_{\rm w} = {\left\langle {\bm u}^2\right\rangle \over \tau_0} \, {\rm Ma}^5 ,
\label{M25}
\end{eqnarray}
where $\tau_0=\ell_0/u_0$ is the turbulent correlation time.
The compressibility contribution $\overline{J}_\nu^{\, ({\rm c})}$
to the rate of the viscous heating is \citep{RK21}
\begin{eqnarray}
\overline{J}_\nu^{\, ({\rm c})} = {\left\langle {\bm u}^2\right\rangle \over \tau_0}
\, {\sigma_c \over 1 + \sigma_c} \, {\rm Re}^{-1/4} .
\label{AA55}
\end{eqnarray}
Equations~(\ref{M25})--(\ref{AA55}) yield the estimate~(\ref{M26}) for the degree of compressibility $\sigma_c$ for small Mach numbers.

\begin{figure}
\centering
\includegraphics[width=7.5cm]{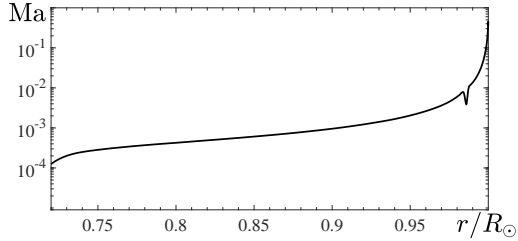}
\caption{\label{Fig4} The radial profile of the Mach number Ma$(r) = u_0/c_{\rm s}$ for the solar convective zone.}
\end{figure}

\begin{figure}
\centering
\includegraphics[width=7.5cm]{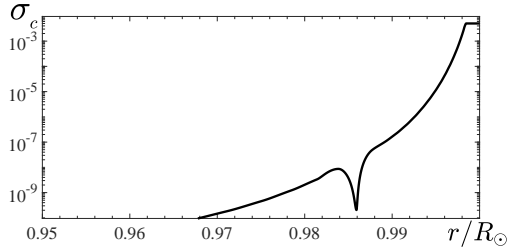}
\caption{\label{Fig5} The radial profile of the degree of compressibility $\sigma_c$ for the solar convective zone.}
\end{figure}

\begin{figure}
\centering
\includegraphics[width=7.5cm]{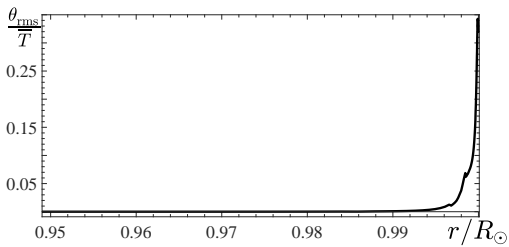}
\caption{\label{Fig6} The radial profile of the r.m.s. of temperature fluctuations
$\theta_{\rm rms}$ measured in the units of the mean temperature
$\meanT$ for the solar convective zone.}
\end{figure}

In Figs.~\ref{Fig4}--\ref{Fig5} we show the radial profiles of the Mach number Ma$(r) = u_0/c_{\rm s}$ and the degree of compressibility $\sigma_c$ of the fluid velocity field
for the solar convective zone based on the model by \cite{SP74}.
The degree of compressibility $\sigma_c$ increases to the surface because the decrease of the sound speed in the vicinity of the solar surface. In Fig.~\ref{Fig6} we plot the radial profile of the r.m.s. of temperature fluctuations $\theta_{\rm rms}$ measured in the units of the mean temperature
$\meanT$ using the above parameters for the solar convective zone. Temperature fluctuations
increase towards to the solar surface due to the compressibility effects.

\begin{figure}
\centering
\includegraphics[width=7.5cm]{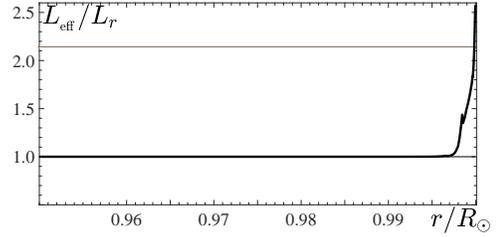}
\caption{\label{Fig7} The radial profile of the ratio $L_{\rm eff} / L_r$  for $a=b=1$ for the solar convective zone.}
\end{figure}

Similar behaviour is observed for fluid density fluctuations $\left\langle \rho'^{\, 2} \right\rangle$ and the temperature-density correlations $\left\langle \theta \, \rho'\right\rangle$ [see equations~(\ref{C8})--(\ref{C9})].
In particular, these second moments enhance towards to the solar surface, resulting in
increase of fluctuations of the radiation absorption coefficient and the effective
penetration length of radiation.
This is seen in Fig.~\ref{Fig7}, where we show the radial profile of the ratio of the turbulence induced effective penetration length of radiation to the mean radiation penetration length $L_{\rm eff} / L_r$ for the solar convective zone based on the model by \cite{SP74}. The ratio $L_{\rm eff} / L_r$ increases in 2.5 times in the vicinity of the solar surface.

\section{Conclusions}

We study a turbulent transport of radiation in the solar convective zone.
To this end, we derive a mean-field equation for the irradiation intensity and show
that due to the turbulent effects the effective penetration length of radiation is increased in several times in comparison with the mean penetration length of radiation that is defined as an inverse mean absorption coefficient.
To demonstrate this effect, we adopt
a model of the solar convective zone based on the mixing length theory.
The mean penetration length of radiation in this model is much smaller than the turbulent integral scale. We have shown that the ratio of the effective penetration length of radiation to the mean penetration length of radiation is increased in 2.5 times in the vicinity of the solar surface.

This effect can be explained by the compressibility effects that become important
in the vicinity of the solar surface, so that
the level of temperature and density fluctuations is increased towards the solar surface.
It causes an increase of fluctuations of the radiation absorption coefficient and the effective penetration length of radiation.
Since the effective penetration length of radiation is changed only in the vicinity of the solar surface (at the depth $\sim 2000$ km), the effect of turbulence on the radiation transport is not strong for the solar type stars. However, this effect can be essential for cold stars (like M3-M5 stars), for which Mach number is larger than that for the sun.

%\section*{Acknowledgments}

\bigskip
\noindent
{\bf Data availability}

\noindent
There are no new data associated with this article.

%\label{lastpage}

\appendix

\section{\bf Temperature and density fluctuations}
\label{Appendix-A}

In this Appendix we derive expression for the level of temperature fluctuations $\left\langle \theta^{2} \right\rangle$, density fluctuations $\left\langle \rho'^{\, 2} \right\rangle$
as well as temperature--density correlations $\left\langle \theta \, \rho'\right\rangle$
using the method described by \cite{RKB18,RK21,RI21}.
The temperature field $T(t,{\bm r})$ in
a compressible fluid velocity field ${\bf U}(t,{\bm r})$ is described by \citep{LL87}
\begin{eqnarray}
{\partial T \over \partial t} + ({\bf U} \cdot {\bm  \nabla})
T + (\gamma - 1) T ({\bm \nabla} \cdot {\bm U})  = D \Delta T + J_\nu ,
\label{APA1}
\end{eqnarray}
where $D$ is the molecular thermal conductivity, $\gamma=c_{\rm p}/c_{\rm v}$ is
the ratio of specific heats and $J_\nu$ is the heating source
due to a viscous dissipation.

We study turbulent flows with large Reynolds $({\rm Re} = u_{0} \, \ell_0 / \nu \gg 1)$
and P\'eclet $({\rm Pe} = u_{0} \, \ell_0 / D \gg 1)$ numbers,
where $u_{0}$ is the characteristic turbulent velocity in the integral
scale $\ell_0$ of turbulence.
Equations for the intensity of temperature fluctuations is derived by means of
the mean-field approach, where the temperature $T= \overline{T} + \theta$,
pressure $P= \overline{P} + p$, density $\rho= \overline{\rho} + \rho'$
and velocity ${\bm U}= \overline{\bm U} + {\bm u}$
are decomposed into mean and fluctuating parts,
with $\overline{T}=\langle T \rangle$ being the mean fluid temperature,
$\overline{P}=\langle P \rangle$ being the mean fluid pressure,
$\overline{\rho}=\langle \rho \rangle$ being the mean fluid density, and
$\overline{\bm U}=\langle {\bm U} \rangle$ being the mean fluid velocity.
Here $\theta$, $p$, $\rho'$ and ${\bm u}$ are fluctuations of temperature,
pressure, density and velocity, respectively, and
the angular brackets denote an ensemble averaging.
Application of the mean-field approach implies
that there is a separation of spatial ($\ell_0 \ll H_T$)
and temporal ($\tau_0 \ll t_T$) scales, where
$H_T$ and $t_T$ are the characteristic spatial and temporal scales
characterizing the variations of the mean temperature field,
and $\tau_0=\ell_0/u_0$.

Ensemble averaging of equation~(\ref{APA1}) yields the mean temperature field:
\begin{eqnarray}
{\partial \overline{T} \over \partial t} + \bec{\nabla} {\bf \cdot}
\left\langle \theta  \, {\bm u} \right\rangle
= - (\gamma - 2) \, \left\langle \theta \, (\bec{\nabla} {\bf \cdot} {\bm u})\right\rangle
+ D \, \Delta \overline{T} + \overline{J}_\nu,
\label{APA2}
\end{eqnarray}
where $\left\langle \theta  \, {\bm u} \right\rangle$ is the turbulent heat flux,
and $\overline{J}_\nu$ is the mean heating source caused
by the viscous dissipation of the turbulent kinetic energy.
Here the case $\overline{\bm U}=0$ is studied for simplicity.
By means of equations~(\ref{APA1}) and~(\ref{APA2}), we obtain equation for
temperature fluctuations, $\theta({\bm x},t)=T-\overline{T}$:
\begin{eqnarray}
{\partial \theta \over \partial t} + {\cal Q} - D \Delta \theta  =
-  ({\bm u}{\bm \cdot} \bec{\nabla}) \overline{T}
- (\gamma - 1) \,\overline{T} \, \bec\nabla {\bm \cdot} \, {\bm u} ,
\label{APA3}
\end{eqnarray}
where ${\cal Q}=\bec\nabla {\bm \cdot} \, (\theta {\bm u}
- \langle {\bm u} \, \theta \rangle)
+ (\gamma - 2) \,[\theta \, \bec\nabla {\bm \cdot} \, {\bm u}
- \langle\theta \, \bec\nabla {\bm \cdot} \, {\bm u}\rangle]$
is the nonlinear term and temperature fluctuations are caused by
the source $-  ({\bm u}{\bm \cdot} \bec{\nabla}) \overline{T}
- (\gamma - 1) \,\overline{T} \, \bec\nabla {\bm \cdot} \, {\bm u}$.
For simplicity we describe the effect of turbulence on the temperature field,
and neglect the feedback effect of the temperature on the turbulence.

We use two-point second-order correlation functions
taking into account small-scale properties of the turbulence, where
the turbulent correlation time and
the turbulent kinetic energy spectrum
are related via the Kolmogorov scalings \citep{MY71,MY75,MC90,F95}.
We adopt the multi-scale approach
\citep{RS75}, and rewrite the two-point second-order correlation
functions as:
\begin{eqnarray}
&& \left\langle \theta({\bm x},t) \, \theta({\bm  y},t)\right\rangle
= \int \,d{\bm k}_1 \, d{\bm k}_2 \left\langle \theta({\bm k}_1,t) \theta({\bm k}_2,t)\right\rangle
\nonumber\\
&& \times \exp\big[i({\bm  k}_1 {\bm \cdot} {\bm x}
+{\bm k}_2 {\bm \cdot}{\bm y})\big] =
\int \Theta^{(II)}({\bm k},{\bm R},t)  \exp[i {\bm k}
{\bm \cdot} {\bm r}] \,d{\bm k} ,
\nonumber\\
\label{SPP1}
\end{eqnarray}
where
\begin{eqnarray}
&& \Theta^{(II)}({\bm k},{\bm R},t) =
 \int \left\langle \theta({\bm k}_1,t) \, \theta({\bm k}_2,t)\right\rangle \, \exp[i {\bm K} {\bm \cdot} {\bm R}] \,d {\bm  K},
\nonumber\\
\label{SAA5}
\end{eqnarray}
and we use large-scale variables: ${\bm R} = ({\bm x}
+ {\bm y}) / 2$, $\, {\bm K} = {\bm k}_1 + {\bm k}_2$,
as well as small-scale variables: ${\bm r} = {\bm x} - {\bm y}$,
$\, {\bm k} = ({\bm k}_1 - {\bm k}_2) / 2$.
Here ${\bm k}_1 = {\bm k} + {\bm  K} / 2$,
and ${\bm k}_2 = - {\bm k} + {\bm  K} / 2$.
Mean-fields depend on the large-scale variables, while
fluctuations depend on the small-scale variables.

The procedure of the derivations of the expressions for the intensity
of temperature fluctuations implies the following steps:
\begin{itemize}
\item{
derivation of equations for the second-order moments in
the Fourier space using the multi-scale approach;}
\item{
application of the spectral $\tau$ approach (see below) which relates the deviations of the third-order moments from those of the background turbulence with the corresponding deviations of the second-order moments;}
\item{
solution of the obtained equations for the second-order moments in the Fourier space;}
\item{
inverse transformation to the physical space to derive expressions for the intensity of temperature fluctuations.}
\end{itemize}

By means of equation~(\ref{APA3}) for temperature fluctuations $\theta$ and the Navier-Stokes equation for velocity fluctuations ${\bm u}$ rewritten in Fourier space, we derive an equation for the second-order moment $\left\langle \theta({\bm k}_1) \, \theta({\bm k}_2) \right\rangle$ as
\begin{eqnarray}
&& {\partial \over \partial t} \left\langle \theta({\bm k}_1) \, \theta({\bm k}_2) \right\rangle=
- \Big[\left\langle u_i({\bm k}_1) \, \theta({\bm k}_2) \right\rangle
\nonumber\\
&& \quad + \left\langle \theta({\bm k}_1) \, u_i({\bm k}_2) \right\rangle\Big]
\, \nabla_i \meanT - (\gamma-1) \, \Big[\left\langle ({\rm div} {\bm u})_{{\bm k}_1} \, \theta({\bm k}_2) \right\rangle
\nonumber\\
&& \quad + \left\langle \theta({\bm k}_1) \, ({\rm div} {\bm u})_{{\bm k}_2} \right\rangle\Big] \meanT
+ \hat{\cal M} \Theta^{(III)},
\label{J10}
\end{eqnarray}
where $\hat{\cal M}\Theta^{(III)}$ are the third-order moment terms caused by the nonlinear terms
in the equation for temperature fluctuations.
Temperature and velocity fluctuations $\theta({\bm k}_{1,2},t)$ and $u_i({\bm k}_{1,2},t)$ depend also on $t$, and the mean temperature $\overline{T}(t,{\bm R})$ depend on $t$ and ${\bm R}$ as well.
For brevity of notations, we do not show these dependencies hereafter.

Equation~(\ref{J10}) for the second-order moments
includes the third-order moments $\hat{\cal M}\Theta^{\rm(III)}$, and
the closure problem arises, i.e.,
how to express the third-order moments $\hat{\cal M} \Theta^{\rm(III)}$
through the lower-order moments \citep{MY71,MY75,MC90}.
We adopt the spectral $\tau$ approach which
postulates that the deviations of the third-moment terms,
$\hat{\cal M} \Theta^{\rm(III)}({\bm k})$, from those afforded by the background
turbulence, $\hat{\cal M} \Theta^{\rm(III,0)}({\bm k})$,
can be expressed through similar deviations
of the second-order moments, $\Theta^{\rm(II)}({\bm k}) -
\Theta^{\rm(II,0)}({\bm k})$ as \citep{O70,PFL76,KRR90}
\begin{eqnarray}
&& \hat{\cal M} \Theta^{\rm(III)}({\bm k}) - \hat{\cal M}
\Theta^{\rm(III,0)}({\bm k})
= - {\Theta^{\rm(II)}({\bm k}) - \Theta^{\rm(II,0)}({\bm k}) \over \tau_r(k)} ,
\nonumber\\
\label{APC4}
\end{eqnarray}
where $\tau_r(k)$ is the scale-dependent relaxation time which can be
identified with the correlation time $\tau(k)$ of the
turbulent velocity field for large Reynolds and P\'eclet numbers.
Since the functions with superscript $(0)$ describe the background
turbulence with a zero turbulent heat flux, equation~(\ref{APC4}) is reduced to
$\hat{\cal M} \Theta^{\rm(III)} = - \left\langle \theta({\bm k}_1)
\, \theta({\bm k}_2) \right\rangle/ \tau(k)$.
We apply the $\tau$ approximation only for
the deviations from the background turbulence, while
the background turbulence is assumed to be known (see below).
Validation of the $\tau$ approximation for different
problems has been performed in various numerical
simulations \citep{BK04,BS05,BSS05,BS05B,BRRK08,RKKB11,RB11,HKRB12,RKB18,EKLR17}.

Since the characteristic times of variation
of the second-order moment $\Theta^{\rm(II)}$ are much
larger than the correlation time $\tau(k)$ in all turbulence scales,
we use the steady-state solution of equation~(\ref{J10}) as
\begin{eqnarray}
&& \left\langle \theta({\bm k}_1) \, \theta({\bm k}_2) \right\rangle =
- \tau(k) \biggl\{\Big[\left\langle u_i({\bm k}_1) \, \theta({\bm k}_2) \right\rangle
\nonumber\\
&& \quad + \left\langle \theta({\bm k}_1) \, u_i({\bm k}_2) \right\rangle\Big] \nabla_i \meanT
+ (\gamma-1) \, \Big[\left\langle ({\rm div} {\bm u})_{{\bm k}_1} \, \theta({\bm k}_2) \right\rangle
\nonumber\\
&& \quad + \left\langle \theta({\bm k}_1) \, ({\rm div} {\bm u})_{{\bm k}_2} \right\rangle\Big] \meanT \biggr\} .
\label{J11}
\end{eqnarray}
Similarly, we derive expression for the second moments entering in equation~(\ref{J11}), i.e., for
$\left\langle u_i({\bm k}_1) \, \theta({\bm k}_2) \right\rangle$  and $\left\langle \theta({\bm k}_1) \, u_j({\bm k}_2) \right\rangle$ as
\begin{eqnarray}
&& \left\langle u_i({\bm k}_1) \, \theta({\bm k}_2) \right\rangle =
- \tau(k) \biggl[\left\langle u_i({\bm k}_1) \, u_j({\bm k}_2) \right\rangle \, \nabla_j \meanT
\nonumber\\
&& \quad+ (\gamma-1) \, \left\langle u_i({\bm k}_1) \, ({\rm div} {\bm u})_{{\bm k}_2} \right\rangle \meanT \biggr],
\label{J12}
\end{eqnarray}
\begin{eqnarray}
&& \left\langle \theta({\bm k}_1) \, u_j({\bm k}_2) \right\rangle =
- \tau(k) \biggl[\left\langle u_i({\bm k}_1) \, u_j({\bm k}_2) \right\rangle \, \nabla_i \meanT
\nonumber\\
&& \quad+ (\gamma-1) \, \left\langle ({\rm div} {\bm u})_{{\bm k}_1} \, u_j({\bm k}_2) \right\rangle \meanT \biggr],
\label{J14}
\end{eqnarray}
and for the second moments $\left\langle ({\rm div} {\bm u})_{{\bm k}_1} \, \theta({\bm k}_2) \right\rangle$ and $\left\langle \theta({\bm k}_1) \,  ({\rm div} {\bm u})_{{\bm k}_2} \right\rangle$:
\begin{eqnarray}
&& \left\langle ({\rm div} {\bm u})_{{\bm k}_1} \, \theta({\bm k}_2) \right\rangle =
- \tau(k) \biggl[\left\langle ({\rm div} {\bm u})_{{\bm k}_1} \, u_j({\bm k}_2) \right\rangle \, \nabla_j \meanT
\nonumber\\
&& \quad + (\gamma-1) \, \left\langle ({\rm div} {\bm u})_{{\bm k}_1} \, ({\rm div} {\bm u})_{{\bm k}_2} \right\rangle \meanT \biggr],
\label{J15}
\end{eqnarray}
\begin{eqnarray}
&& \left\langle \theta({\bm k}_1) \,  ({\rm div} {\bm u})_{{\bm k}_2} \right\rangle =
- \tau(k) \biggl[\left\langle u_i({\bm k}_1) \, ({\rm div} {\bm u})_{{\bm k}_2}  \right\rangle \, \nabla_i \meanT
\nonumber\\
&& \quad+ (\gamma-1) \, \left\langle ({\rm div} {\bm u})_{{\bm k}_1} \, ({\rm div} {\bm u})_{{\bm k}_2} \right\rangle \meanT \biggr] .
\label{J16}
\end{eqnarray}
Substituting equations~(\ref{J12})--(\ref{J16}) into equation~(\ref{J11}),
we obtain
\begin{eqnarray}
&& \left\langle \theta({\bm k}_1) \, \theta({\bm k}_2) \right\rangle =
2\tau^2(k) \biggl\{\left\langle u_i({\bm k}_1) \, u_j({\bm k}_2) \right\rangle \, (\nabla_i \meanT)
(\nabla_j \meanT)
\nonumber\\
&& + (\gamma-1) \, \Big[\left\langle ({\rm div} {\bm u})_{{\bm k}_1} \, u_j({\bm k}_2) \right\rangle
+ \left\langle u_j({\bm k}_1) \, ({\rm div} {\bm u})_{{\bm k}_2} \right\rangle\Big]
\nonumber\\
&& \times \meanT \, \nabla_j\meanT + (\gamma-1)^2 \, \left\langle ({\rm div} {\bm u})_{{\bm k}_1} \, ({\rm div} {\bm u})_{{\bm k}_2} \right\rangle \meanT^2 \biggr\} .
\label{J17}
\end{eqnarray}

Since all terms in equations~(\ref{J11}) and~(\ref{J17}) are proportional to either $({\bm \nabla} \meanT)^2$, or $({\bm \nabla} \meanrho)^2$, or $({\bm \nabla} \meanT) ({\bm \nabla} \meanrho)$ (see below), and we consider homogeneous
density stratified turbulence, we do not need to perform additional Taylor expansions
over small parameters $\ell_0/H_T$ and $\ell_0/H_\rho$ in these
terms \citep{RK21,RI21}, where $H_T$ and $H_\rho$ are the characteristic scales of variations of the mean temperature and mean density, respectively. In particular, hereafter we neglect small terms $\sim {\rm O}[(\ell_0/H_T)^3, (\ell_0/H_\rho)^3]$.
This implies that we replace ${\bm k}_1$ by ${\bm k}$ and ${\bm k}_2$
by $-{\bm k}$ in all second moments in equation~(\ref{J17}).

In equation~(\ref{J17}) we take into account a one way coupling,
i.e., we neglect the feedback effect of the mean temperature gradients
on the turbulent velocity field.
This implies that we replace the correlation function
$f_{ij}=\left\langle u_i({\bm k}) \, u_j(-{\bm k}) \right\rangle$ in equation~(\ref{J17}) by $f_{ij}^{(0)}
=\left\langle u_i({\bm k}) \, u_j(-{\bm k}) \right\rangle^{(0)}$
for the background turbulence with a zero turbulent heat flux.
Similarly we replace $\left\langle ({\rm div} {\bm u})_{{\bm k}} \, u_j(-{\bm k}) \right\rangle$,
$\left\langle u_j({\bm k}) \, ({\rm div} {\bm u})_{-{\bm k}} \right\rangle$
and $\left\langle ({\rm div} {\bm u})_{{\bm k}} \, ({\rm div} {\bm u})_{-{\bm k}} \right\rangle$
in equation~(\ref{J17}) by the corresponding correlation functions for the background turbulence with a zero turbulent heat flux.

To find the intensity of temperature fluctuations $\left\langle  \theta^2 \right\rangle$ for large P\'eclet numbers, we adopt a model for the background turbulence, $f_{ij}^{(0)}({\bm k})=\left\langle u_i({\bm k}) \, u_j(-{\bm k}) \right\rangle^{(0)}$, that is
a statistically stationary density-stratified
compressible turbulence given by \citep{EKR95,EKR17,RI21}:
\begin{eqnarray}
&& f_{ij}^{(0)}({\bm k}) = {1 \over 8 \pi \, k^2 \,  (1+ \sigma_c)} \,
\biggl\{E(k)\, \biggl[(\delta_{ij} - k_{ij}) \, \left(1 - {\lambda^2 \over k^2}\right)
\nonumber\\
&& \quad + {\lambda^2 \over k^2} \, \big(\delta_{ij} - \lambda_{ij}\big)\biggr]
+ {{\rm i} \over k^2} \, \left[E(k) + \sigma_c \, E_c(k)\right] \,  \Big(k_j \lambda_i
\nonumber\\
&& \quad - k_i \lambda_j\Big) + 2 \sigma_c \, E_c(k) \, k_{ij}\biggr\} \left\langle {\bm u}^2\right\rangle ,
\label{J18}
\end{eqnarray}
where $k_{ij}=k_{i} \, k_{j}/k^2$, $\lambda_{ij}=\lambda_{i} \, \lambda_{j}/\lambda^2$,
and ${\bm \lambda} = - {\bm \nabla} \ln \meanrho$ characterizes the
fluid density stratification.
This model is different from that derived by
\cite{RKB18,RK21}.
In particular, this model takes into account a strong density stratification.
In addition, the turbulent flux $\langle \rho' {\bm u}\rangle$
is very small $[\sim {\rm O}(\lambda \, \ell_0)^3]$.

The background turbulence is of Kolmogorov type with
a constant energy flux over the spectrum,
i.e., the turbulent kinetic energy spectrum for the incompressible part
of turbulence in the inertial range $k_0<k<k_\nu$ is
$E(k) = - d \tilde \tau(k) / dk$. Here $\tilde \tau(k) =
(k / k_{0})^{1-q}$ with $1 < q < 3$ is the
exponent of the turbulent kinetic energy spectrum.
Similarly, the turbulent kinetic energy spectrum
for the compressible part of turbulence is $E_c(k) = - d \tilde \tau_c(k) / dk$,
where $\tilde \tau_c(k) =(k / k_{0})^{1-q_c}$ with $1 < q_c < 3$.
For example, the exponent of the incompressible part of
the spectrum, $q=5/3$, corresponds to the Kolmogorov spectrum,
while the exponent of the compressible part of
the spectrum, $q_c=2$, describes the Burgers
turbulence with shock waves. These exponents of the spectra are observed
in numerical simulations in compressible turbulence \citep{KN07,FE13}.
The correlation time for a compressible turbulence in the Forier space is \citep{RK21}
\begin{eqnarray}
\tau(k) = {2\tau_0 \over 1+\sigma_c} \, \Big[\tilde\tau(k) + \sigma_c \,\tilde \tau_c(k)\Big] .
\label{J19}
\end{eqnarray}

To determine the level of temperature fluctuations $\left\langle  \theta^2 \right\rangle
= \int \tau^2(k) \left\langle \theta({\bm k}) \, \theta(-{\bm k}) \right\rangle \,d{\bm k}$ for large P\'eclet numbers, we use equations~(\ref{J17})--(\ref{J19}).
To this end, we calculate the following integrals:
\begin{eqnarray}
\int \tau^2(k) \left\langle u_i({\bm k}) \, u_j(-{\bm k}) \right\rangle^{(0)} \,d{\bm k} =
{4 \over 9} \, \ell_0^2 \, \delta_{ij} ,
\label{J20}
\end{eqnarray}

\begin{eqnarray}
\int \tau^2(k) \left\langle u_i({\bm k}) \, ({\rm div} {\bm u})_{-{\bm k}} \right\rangle^{(0)} \,d{\bm k} = {4 \over 9} \, \ell_0^2 \, \lambda_{i} ,
\label{J21}
\end{eqnarray}

\begin{eqnarray}
&&\int \tau^2(k) \left\langle ({\rm div} {\bm u})_{{\bm k}} \, ({\rm div} {\bm u})_{-{\bm k}} \right\rangle^{(0)} \,d{\bm k} = {4 \over 9} \, (\ell_0 \, \lambda)^2
\nonumber\\
&& \quad + 4 f_c \, \left({\sigma_c \over 1 + \sigma_c} \right)^3 \, \left[1 - {1 \over 9} \, (\ell_0 \, \lambda)^2\right] .
\label{J22}
\end{eqnarray}
For the integration in ${\bm k}$ space in equations~(\ref{J20})--(\ref{J22}),
we use the following identities:
\begin{eqnarray}
\int_{k_0}^{k_\nu} \tau^2(k) \, \left[E(k) + \sigma_c \, E_c(k)\right] \, dk = {4 \over 3} \, \tau_0^2 \, (1 + \sigma_c) ,
\label{I2}
\end{eqnarray}

\begin{eqnarray}
\int_{k_0}^{k_\nu} \tau^2(k) \, k^2 \, E_c(k) \, dk = 4 f_c \, \left({\tau_0 \over \ell_0}\right)^2 \,  \left({\sigma_c \over 1 + \sigma_c} \right)^2 .
\label{I1}
\end{eqnarray}
Therefore, equations~(\ref{J17}) and~(\ref{J20})--(\ref{J22}) yield the level of temperature fluctuations $\left\langle  \theta^2 \right\rangle$ for large P\'eclet numbers given by equation~(\ref{DD1}).

Derivation of equations~(\ref{C8}) and~(\ref{C9}) for $\left\langle \rho'^{\, 2} \right\rangle$
and $\left\langle \theta \, \rho'\right\rangle$ is performed in a similar way. In particular, using the continuity equation for the fluid density fluctuations $\rho'$ written in the Fourier space, we obtain the evolutionary equation for the second moment $\left\langle \rho'({\bm k}_1) \, \rho'({\bm k}_2) \right\rangle$ as
\begin{eqnarray}
&& {\partial \over \partial t} \left\langle \rho'({\bm k}_1) \, \rho'({\bm k}_2) \right\rangle=
- \Big[\left\langle u_i({\bm k}_1) \, \rho'({\bm k}_2) \right\rangle
\nonumber\\
&& \quad + \left\langle \rho'({\bm k}_1) \, u_i({\bm k}_2) \right\rangle\Big]
\, \nabla_i \meanrho - \Big[\left\langle ({\rm div} {\bm u})_{{\bm k}_1} \, \rho({\bm k}_2) \right\rangle
\nonumber\\
&& \quad + \left\langle \rho({\bm k}_1) \, ({\rm div} {\bm u})_{{\bm k}_2} \right\rangle\Big] \meanrho + \hat{\cal M}\rho^{(III)},
\label{I10}
\end{eqnarray}
where $\hat{\cal M}\rho^{(III)}$ are the third-order moment terms
related to nonlinear terms in the equation for density fluctuations.

Applying the spectral $\tau$ approach,
we obtain expression for the second moment $\left\langle \rho'({\bm k}_1) \, \rho'({\bm k}_2) \right\rangle$ as
\begin{eqnarray}
&& \left\langle \rho'({\bm k}_1) \, \rho'({\bm k}_2) \right\rangle =
- \tau(k) \biggl\{\Big[\left\langle u_i({\bm k}_1) \, \rho'({\bm k}_2) \right\rangle
\nonumber\\
&& \quad + \left\langle \rho'({\bm k}_1) \, u_i({\bm k}_2) \right\rangle\Big] \nabla_i \meanrho
+ \Big[\left\langle ({\rm div} {\bm u})_{{\bm k}_1} \, \rho'({\bm k}_2) \right\rangle
\nonumber\\
&& \quad + \left\langle \rho'({\bm k}_1) \, ({\rm div} {\bm u})_{{\bm k}_2} \right\rangle\Big] \meanrho \biggr\} .
\label{I11}
\end{eqnarray}
Similarly, we derive expression for the second moments $\left\langle u_i({\bm k}_1) \, \rho'({\bm k}_2) \right\rangle$  and
$\left\langle \rho'({\bm k}_1) \, u_j({\bm k}_2) \right\rangle$ as
\begin{eqnarray}
\left\langle u_i({\bm k}_1) \, \rho'({\bm k}_2) \right\rangle &=&
- \tau(k) \biggl[\left\langle u_i({\bm k}_1) \, u_j({\bm k}_2) \right\rangle \, \nabla_j \meanrho
\nonumber\\
&& \quad + \left\langle u_i({\bm k}_1) \, ({\rm div} {\bm u})_{{\bm k}_2} \right\rangle \meanrho \biggr],
\label{I12}
\end{eqnarray}
\begin{eqnarray}
\left\langle \rho'({\bm k}_1) \, u_j({\bm k}_2) \right\rangle &=&
- \tau(k) \biggl[\left\langle u_i({\bm k}_1) \, u_j({\bm k}_2) \right\rangle \, \nabla_i \meanrho
\nonumber\\
&& + \left\langle ({\rm div} {\bm u})_{{\bm k}_1} \, u_j({\bm k}_2) \right\rangle \meanrho \biggr],
\label{I14}
\end{eqnarray}
and for $\left\langle ({\rm div} {\bm u})_{{\bm k}_1} \, \rho'({\bm k}_2) \right\rangle$ and $\left\langle \rho'({\bm k}_1) \,  ({\rm div} {\bm u})_{{\bm k}_2} \right\rangle$:
\begin{eqnarray}
\left\langle ({\rm div} {\bm u})_{{\bm k}_1} \, \rho'({\bm k}_2) \right\rangle &=&
- \tau(k) \biggl[\left\langle ({\rm div} {\bm u})_{{\bm k}_1} \, u_j({\bm k}_2) \right\rangle \, \nabla_j \meanrho
\nonumber\\
&& + \left\langle ({\rm div} {\bm u})_{{\bm k}_1} \, ({\rm div} {\bm u})_{{\bm k}_2} \right\rangle \meanrho \biggr],
\label{I15}
\end{eqnarray}
\begin{eqnarray}
\left\langle \rho'({\bm k}_1) \,  ({\rm div} {\bm u})_{{\bm k}_2} \right\rangle &=&
- \tau(k) \biggl[\left\langle u_i({\bm k}_1) \, ({\rm div} {\bm u})_{{\bm k}_2}  \right\rangle \, \nabla_i \meanrho
\nonumber\\
&& + \left\langle ({\rm div} {\bm u})_{{\bm k}_1} \, ({\rm div} {\bm u})_{{\bm k}_2} \right\rangle \meanrho \biggr] .
\label{I16}
\end{eqnarray}
Substituting equations~(\ref{I12})--(\ref{I16}) into equation~(\ref{I11})
we obtain
\begin{eqnarray}
&& \left\langle \rho'({\bm k}_1) \, \rho'({\bm k}_2) \right\rangle =
2\tau^2(k) \biggl\{\left\langle u_i({\bm k}_1) \, u_j({\bm k}_2) \right\rangle \, (\nabla_i \meanrho)
(\nabla_j \meanrho)
\nonumber\\
&& \quad + \Big[\left\langle ({\rm div} {\bm u})_{{\bm k}_1} \, u_j({\bm k}_2) \right\rangle
+ \left\langle u_j({\bm k}_1) \, ({\rm div} {\bm u})_{{\bm k}_2} \right\rangle\Big] \meanrho \, \nabla_j\meanrho
\nonumber\\
&& \quad + \left\langle ({\rm div} {\bm u})_{{\bm k}_1} \, ({\rm div} {\bm u})_{{\bm k}_2} \right\rangle \meanrho^2 \biggr\} .
\label{I17}
\end{eqnarray}
We do not take into account small terms $\sim {\rm O}[(\ell_0/H_\rho)^3]$ in equations~(\ref{I11}) and~(\ref{I17}). Since all terms in equations~(\ref{I11}) and~(\ref{I17}) are proportional to $({\bm \nabla} \meanrho)^2$ and we consider homogeneous density stratified turbulence, we do not need to perform additional Taylor expansions over small parameter $\ell_0/H_\rho$ in these
terms. Thus, we replace ${\bm k}_1$ by ${\bm k}$ and ${\bm k}_2$
by $-{\bm k}$ in all second moments in equation~(\ref{I17}).
Using equations~(\ref{J20})--(\ref{J22}) and~(\ref{I17}), we determine the level of density fluctuations $\left\langle \rho'^{\, 2} \right\rangle= \int \tau^2(k) \left\langle \rho'({\bm k}) \, \rho'(-{\bm k}) \right\rangle \,d{\bm k}$ for large P\'eclet numbers given by equation~(\ref{C8}).

Now we derive the evolutionary equation for the second moment $\left\langle \theta({\bm k}_1) \, \rho'({\bm k}_2) \right\rangle$ using the continuity equation for the fluid density fluctuations and the equation for the temperature fluctuations written in the Fourier space:
\begin{eqnarray}
&& {\partial \over \partial t} \left\langle \theta({\bm k}_1) \, \rho'({\bm k}_2) \right\rangle=
- \left\langle u_i({\bm k}_1) \, \rho'({\bm k}_2) \right\rangle \, \nabla_i \meanT
\nonumber\\
&& \quad - \left\langle \theta({\bm k}_1) \, u_j({\bm k}_2) \right\rangle \nabla_j \meanrho
- (\gamma-1) \, \left\langle ({\rm div} {\bm u})_{{\bm k}_1} \, \rho'({\bm k}_2) \right\rangle \, \meanT
\nonumber\\
&& \quad - \left\langle \theta({\bm k}_1) \, ({\rm div} {\bm u})_{{\bm k}_2} \right\rangle \meanrho
+ \hat{\cal M}\Theta_\rho^{(III)},
\label{I18}
\end{eqnarray}
where $\hat{\cal M}\Theta_\rho^{(III)}$ are the third-order moment terms related to the nonlinear terms
in the equations for temperature and density fluctuations. Applying the spectral $\tau$ approach,
we obtain expression for the second moment $\left\langle \theta({\bm k}_1) \, \rho'({\bm k}_2) \right\rangle$ as
\begin{eqnarray}
&& \left\langle \theta({\bm k}_1) \, \rho'({\bm k}_2) \right\rangle =
- \tau(k) \biggl[\left\langle u_i({\bm k}_1) \, \rho'({\bm k}_2) \right\rangle \, \nabla_i \meanT
\nonumber\\
&& \quad + \left\langle \theta({\bm k}_1) \, u_j({\bm k}_2) \right\rangle \nabla_j \meanrho
+ (\gamma-1) \, \left\langle ({\rm div} {\bm u})_{{\bm k}_1} \, \rho'({\bm k}_2) \right\rangle \, \meanT
\nonumber\\
&& \quad + \left\langle \theta({\bm k}_1) \, ({\rm div} {\bm u})_{{\bm k}_2} \right\rangle \meanrho \biggr] .
\label{I19}
\end{eqnarray}
Substituting equations~(\ref{J14}), (\ref{J16}), (\ref{I12}), (\ref{I15}) into equation~(\ref{I19}),
we obtain
\begin{eqnarray}
&& \left\langle \theta({\bm k}_1) \, \rho'({\bm k}_2) \right\rangle = 2 \tau^2(k) \biggl\{\left\langle u_i({\bm k}_1) \, u_j({\bm k}_2) \right\rangle \, (\nabla_i \meanT) \, (\nabla_j \meanrho)
\nonumber\\
&& + (\gamma-1) \, \Big[\left\langle ({\rm div} {\bm u})_{{\bm k}_1} \, ({\rm div} {\bm u})_{{\bm k}_2} \right\rangle \,\meanrho
+ \left\langle ({\rm div} {\bm u})_{{\bm k}_1} \, u_j({\bm k}_2) \right\rangle
\nonumber\\
&& \times (\nabla_j \meanrho) \Big] \, \meanT
+ \left\langle u_i({\bm k}_1) \, ({\rm div} {\bm u})_{{\bm k}_2} \right\rangle\, \meanrho \, (\nabla_i\meanT) \biggr\} .
\label{I24}
\end{eqnarray}
We neglect small terms $\sim {\rm O}[(\ell_0/H_T)^3, (\ell_0/H_\rho)^3]$ in equations~(\ref{I19}) and~(\ref{I24}).
Since all terms in equations~(\ref{I19}) and~(\ref{I24}) are proportional to either $({\bm \nabla} \meanrho)^2$ or $({\bm \nabla} \meanT) ({\bm \nabla} \meanrho)$, and we consider homogeneous density stratified turbulence, we do not need to perform additional Taylor expansions
over small parameters $\ell_0/H_T$ and $\ell_0/H_\rho$ in these
terms. Therefore, we replace ${\bm k}_1$ by ${\bm k}$ and ${\bm k}_2$
by $-{\bm k}$ in all second moments in equation~(\ref{I24}).
Using equations~(\ref{J20})--(\ref{J22}) and~(\ref{I24}), we determine  $\left\langle \theta \, \rho'\right\rangle = \int \tau^2(k) \left\langle \theta({\bm k}) \, \rho'(-{\bm k}) \right\rangle \,d{\bm k}$ for large P\'eclet numbers given by equation~(\ref{C9}).

\end{document}